\documentclass[aip,jmp,amsmath,amssymb,reprint]{revtex4-1}
\usepackage[T1]{fontenc}
\usepackage[latin9]{inputenc}
\usepackage[letterpaper]{geometry}
\usepackage{textcomp}
\usepackage{amsmath}
\usepackage{graphicx}
\usepackage{amssymb}

\begin{document}

\title{{\normalsize Suppression of decoherence in a graphene monolayer ring}}
\author{{\normalsize D. Smirnov}}
\email{smirnov@nano.uni-hannover.de}
\author{{\normalsize J. C. Rode}}
\author{{\normalsize R. J. Haug}}
\address{Institut f\"ur Festk\"orperphysik, Leibniz Universit\"at Hannover, Appelstr.
2 30167 Hannover, Germany}

\begin{abstract}
The influence of high magnetic fields on coherent transport is investigated. A monolayer graphene quantum ring is fabricated and the Aharonov-Bohm effect is observed. For increased magnitude of the magnetic field higher harmonics appear. This phenomenon is attributed to an increase of the phase coherence length due to reduction of spin flip scattering.  
\end{abstract}

\maketitle

The Aharonov-Bohm (AB) effect \cite{AharonovBohmEffektFP1959[5]} is one of the most prominent effects to directly observe quantum interference. It was studied in numerous publications over the last years in ring shaped metals\cite{Ahronov-Bohm_firstExperiment} and semiconducting heterostructures\cite{qdring1,qdring2,qdring3}. Higher harmonics of the AB effect appear when the charge carriers interfere after passing the ring more than once. However the $N^{th}$ harmonic only occurs if the most decisive factor, the phase coherence length $l_{\phi}$, is in the order of $N+1$ times the system length, i.e. ring circumference. Recently the AB effect gained interest in experimental\cite{AB-Exp-1,AB-Exp-2,MyAAB-Exp-3,Markovic-AB-Expr-4,AB-Mirrors-Expr-5} and theoretical\cite{AB-Theory1,AB-Theory2,TheorieKleintunneling[13],AB-Theory3} studies for monolayer graphene\cite{Erscheinungspaper2004[1]}. Although higher harmonics were already reported for most of the semiconducting heterostructures, only in two publications higher harmonics are mentioned for graphene structures \cite{AB-Exp-1,AB-Mirrors-Expr-5}. However in both studies specific circumstances were established: Observation of the first harmonic only at specific magnetic fields\cite{AB-Exp-1} or higher harmonics with superconducting mirrors at the leads of the ring\cite{AB-Mirrors-Expr-5}. In order to observe higher AB harmonics the phase coherence length has to increase. Therefore ways to suppress some of the decoherence mechanisms have to be further investigated.

\begin{figure}
\begin{centering}
\includegraphics{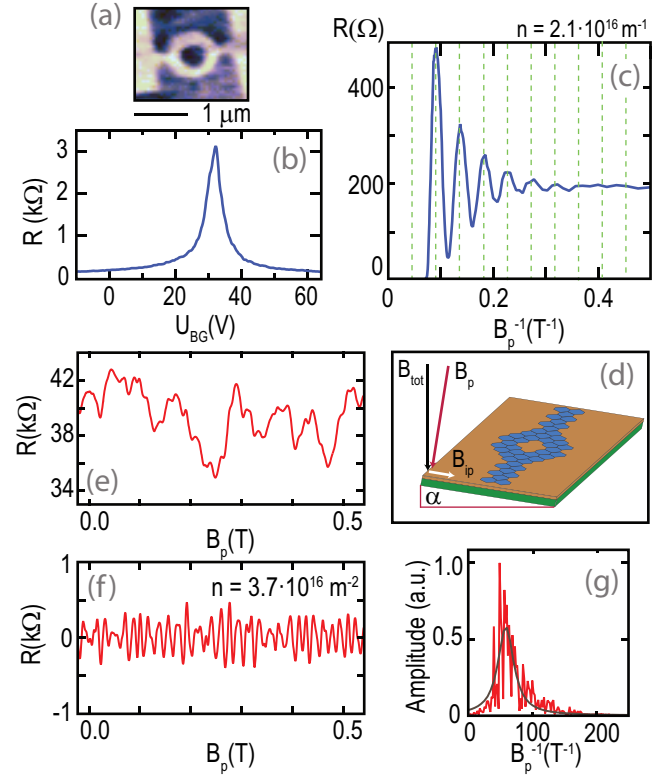}

\end{centering}
\caption{\label{Fig1} {(a)~AFM picture of the measured graphene ring sample. (b)~Four probe characterization measurement for an area near the ring versus backgate (BG) voltage. (c)~Shubnikov-de Haas oscillations characterizing the sample as monolayer graphene. (d)~Schematic picture of the graphene ring with a tilt angle $\mathrm{\alpha}$. (e)~Aharonov-Bohm resistance measurements. (f)~The filtered oscillation from the same measurement. (g)~The Fourier spectrum after filtering (red) with a Lorentzian fit (black).}}
\end{figure}
 
In this letter a monolayer graphene ring is investigated. Our experimental setup allowed to vary the perpendicular and the in-plane magnetic field. The AB effect is studied in such conditions and the impact of magnetic field and its components is analyzed.  

The sample was fabricated using standard procedures: Graphene was placed on $330\,\mathrm{nm}\ \mathrm{Si/SiO_{2}}$ substrate via scotch tape method. The flake was identified as a monolayer by optical microscopy using the contrast shift in the red channel\cite{MakingGrapheneVisible}. Afterwards a ring was formed via electron beam lithography and plasma oxygen etching with an average radius of $290\,\mathrm{nm}$ and a width of $200\,\mathrm{nm}$. Figure~1(a) shows an atomic force microscope (AFM) picture of the device. Before the evaporation of Chromium/Gold contacts the sample was cleaned by AFM contact mode scanning \cite{AFMCleaning} to increase overall quality. The as prepared device was then loaded in a $\mathrm{He^{3}/He^{4}}$ mixing chamber with a base temperature of $100\,\mathrm{mK}$ and a total magnetic field $B_{tot}$ up to $14\,\mathrm{T}$. Furthermore the experimental setup allowed tilting of the sample in respect to $B_{tot}$ (Fig.~1(d)) leading to perpendicular magnetic field $B_{p}$ and in-plane magnetic field $B_{ip}$ components. The resistance was measured with a lock-in amplifier with a current of $5\,\mathrm{nA}$.

The characterization of the sample was performed in a four-terminal setup using contacts near the ring. Figure~1(b) shows the resistance measurement versus the backgate voltage $U_{BG}$. The charge neutrality point (CNP) can be identified at $U_{BG}=32\,V$. This doping  is attributed to the fabrication process.  From the four-terminal measurements the mobility  $\mathrm{\mu = 18.000\,cm^2/Vs}$ for electrons and $\mathrm{\mu = 20.000\,cm^2/Vs}$ for holes was calculated. This can be attributed to the used type of cleaning, which increases the mobility but doesn't reduce doping in the same way. Figure 1(c) shows the longitudinal magnetoresistance versus the inverse magnetic field. Shubnikov-de Haas oscillations are visible with a Berry's Phase of $\pi$ which identifies the flake as monolayer \cite{Berry Phase1,Berry Phase2}. 

Figure~1(e) shows AB measurements of the ring for a fixed charge carrier concentration of $n_{e}=3.7\cdot 10^{16}\,\mathrm{m^{-2}}$. Due to a limited number of contacts on one side of the ring a three-terminal setup was used. Oscillations with an average visibility of $\mathrm{1.5\,\%}$ on top of the background signal are observed. These oscillations can be identified as AB effect. The period of these oscillations has a mean value of $\Delta B\mathrm{=16.7\,mT}$ for the shown data. Overall, the average AB period for different measurements is $\Delta B=16.0\pm 1.0\,\mathrm{mT}$. This corresponds to a ring radius $\mathrm{287.5\pm 7.5\,nm}$ which fits the geometry of the ring well. The background fluctuations can be identified as universal conductance fluctuations. To separate the AB oscillations from the background fluctuations the fast Fourier transform of the measured signal is multiplied with a smooth high-pass filter. The filtered AB oscillations are shown in Fig.~1(f). Figure~1(g) shows a Lorentzian fit to the Fourier spectrum. The original AB oscillations are visible, however further contributions at higher frequencies are not observed. 

From the ring conductance one can estimate the diffusion constant using the Einstein relation $\sigma=\nu^2eD$ with the Density of States $\nu=4\sqrt{n\pi}/h v_F$ , $v_F$ the Fermi velocity, to $D\approx\mathrm{0.01\,m^2/s}$. By estimating $\tau_{\phi}$ from\cite{Weak_localization_Estimation} ($\tau_\phi ^{-1}\approx k_BT\cdot \mathrm{ln}(g)/(\hbar g)$, $g=\sigma h/e^2$) we can calculate the phase coherence length $l_{\phi}=\sqrt{D\tau_{\phi}}$ to $l_{\phi}\approx 1.6\,\mathrm{\mu m}$. This estimation is in agreement with our observation. The phase coherence length $l_{\phi}$ is in the range of the ring circumference $L_C=\mathrm{1.8\,\mu m}$ which explains the observation of the original AB oscillations and the absence of higher harmonics. The low visibility of the oscillations might be resulting not only from the low phase coherence length, but also from the high number of modes in our system. However, it is comparable with AB oscillations shown in monolayer graphene in previous experiments\cite{AB-Exp-1,AB-Exp-2,MyAAB-Exp-3,Markovic-AB-Expr-4,AB-Mirrors-Expr-5} even for much lower charge carrier concentrations.

Figure~2(a) shows the AB measurements for a fixed charge carrier concentration of $n_{e}=2.1\cdot 10^{16} \mathrm{m^{-2}}$. The sample was tilted by $25^\circ$ and the range of the total magnetic field is increased so the influence of a high magnetic field and its components can be studied. Nor Shubnikov-de Haas oscillations nor the Quantum Hall effect are observed in the magnetotransport measurement, yet the AB effect is clearly visible even for high magnetic fields (Fig.~2(b)-(d)). The absence of Landau level quantization indicates that the ring does not achieve the same quality as the rest of the flake, presumably caused by strong edge disorder. To analyze the influence of high magnetic fields on the AB effect the measurement was split up into different parts. The range of these regions (Fig.~2a) was chosen to be small enough to clearly observe the influence of $B_{tot}$ or its components but high enough to resolve clear features in the Fourier spectrum. 

\begin{figure}
\begin{centering}
\includegraphics{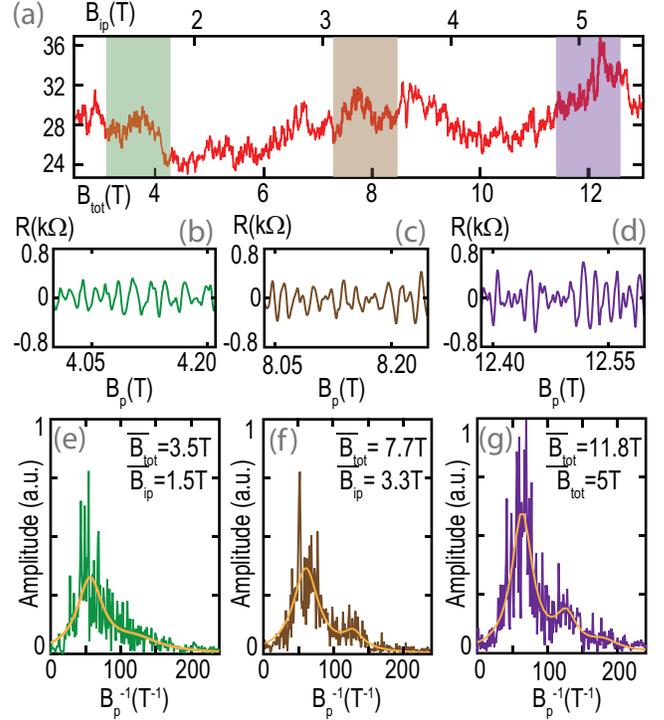}
\end{centering}
\caption{\label{Fig2} {(a)~Magnetotransport measurements for a tilt angle $\alpha=25^\circ$. Three regions are marked representing $\overline{B}_{tot}=3.5\,\mathrm{T}$, $7.7\,\mathrm{T}$, and $11.8\,\mathrm{T}$. (b)-(d)~Filtered oscillations (as described in text). (e)-(g)~Filtered Fourier spectra for the marked regions. A development of a second peak is clearly visible.}}
\end{figure}

Figure~2(e)-(g) shows equally scaled Fourier spectra centred around the marked regions in Fig.~2(a). The Fourier spectrum for $\overline{B}_{tot}=3.5\,\mathrm{T}$ shows a clear developed first peak corresponding to the original AB oscillations with a shoulder at higher frequencies next to it. For higher magnetic fields the shoulder develops into a clearly visible second peak (Fig.~2(f)). The spectra were fitted with a double Lorentzian fit. The average position of the second peak is $B_p^{-1}=\mathrm{115 \pm 10\,T^{-1}}$ which corresponds to a period of $\Delta B=8.75\pm 0.75\,\mathrm{mT}$. Increasing the magnetic field even further not only leads to a more visible second peak but a higher first peak as well (Fig.~2(g)). This corresponds to an amplitude gain for the original AB oscillations (Fig.~2(d)). The second peak can be explained as the first harmonic of the AB oscillations and fits the geometry of the ring with an expected period of $\Delta B_{g}\sim 8\,\mathrm{mT}$ quite well. Additionally to the observed first harmonic a shoulder-like feature is visible at $\overline{B}_{tot}=11.8\,\mathrm{T}$. It was analyzed with a further Lorentzian fit and the resulting period is $\Delta B\approx 5.5\,\mathrm{mT}$. The expected period for the second AB harmonic is $\Delta B\approx 6\,\mathrm{mT}$, therefore the observed shoulder is identified as the second AB harmonic.

Due to the observation of the first and second AB harmonic in Fig.~2(g) one can estimate the coherence length to be in between two to three times the ring circumference at $\overline{B}_{tot}=11.8\,\mathrm{T}$, i.e. $l_{\phi}=4.5\,\mathrm{\mu m}$. So by increasing the magnitude of the magnetic field the phase coherence length seems to increase as well. With high enough  magnetic field the phase coherence length comes in range of two times the ring circumference and the first harmonic becomes visible (Fig.~2(e),(f)). A longer phase coherence length also leads to a full development of the original AB oscillation which is observed in the development of the main peak in Fig.~2(e)-(g). The phase coherence length can be influenced by a number of scattering processes. Due to the used preparation methods a strong disorder at the edges is expected, leading to charge traps or defects. Furthermore an evenly distributed presence of defects in the $\mathrm{SiO_2}$ substrate and adatoms on top of the flake is certain due to the observed doping (Fig. 1(b)). However the absence of Landau level quantization over the ring in comparison to a clear observation next to it (Fig.~1,2) is evident for a higher impact of edge disorder on the transport through the ring. Recently it was shown\cite{Kats1,Kats2} that fluctuations of the magnetic moments can occur at the edges of graphene nanoribbons resulting in spin-flip scattering. Such spin-flip scattering will lead to decoherence and a suppression of the phase coherence. Applying a magnetic field prevents this decoherence mechanism by spin polarisation. As a result the phase coherence length is increasing and leads to the observed phenomenon. We can estimate the decoherence rate reduction from the observed increase of the phase coherence length from $l_{\phi,B_{tot}\approx 0\,\mathrm{T}}=1.6\,\mathrm{\mu m}$ to $l_{\phi,B_{tot}\approx 11.8\,\mathrm{T}}=4.5\,\mathrm{\mu m}$, and the obtained value is $\tau_{diff}^{-1}\approx 3.6\,\mathrm{ns}^{-1}$. An increase of the phase coherence length with magnetic fields was also shown in experiments studying universal conductance fluctuations in in-plane magnetic field\cite{UCF_in-plane magnetic field} and was attributed to interaction with magnetic defects in the substrate. Although our sample geometry differs strongly from\cite{UCF_in-plane magnetic field}, the resulting observation and decoherence rate reduction are comparable, which is a further support of the influence of spin polarisation.

\begin{figure}
\begin{centering}
\includegraphics{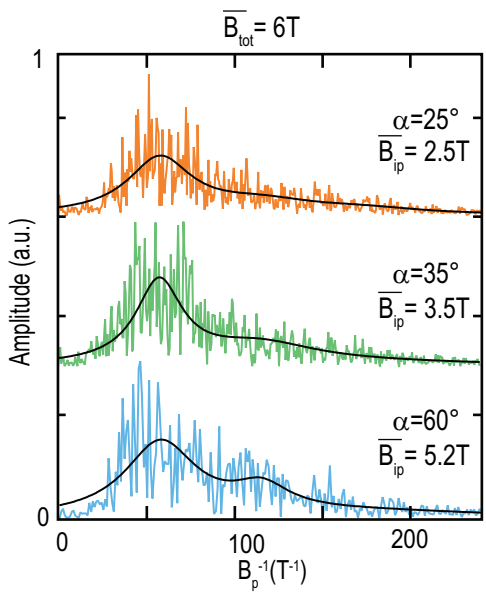}
\end{centering}
\caption{\label{Fig3} {Fourier spectra for AB measurement with different tilt angles $\alpha=25^\circ$, $35^\circ$, and $60^\circ$. The average of the total magnetic field is kept constant at $B_{tot}=6\,\mathrm{T}$  }}
\end{figure}

In Fig.~2 perpendicular and parallel components of the magnetic field are variated, so the impact of only the in-plane magnetic field is not clear. Figure 3 shows the Fourier spectra for AB measurements with different tilt angles. Though the mean magnitude of the total magnetic field is kept constant the in-plane component increases with higher tilt angles. With rising tilt angles a development of the first harmonic is observed. The peaks are analyzed as described before and show the same result for their positions. As a result one can conclude that the in-plane component alone seems to increase the phase coherence length at constant total magnetic field. However the main effect seems to stem from the total magnetic field.
Although the in-plane magnetic fields for $\alpha=35^\circ$ and $\alpha=60^\circ$ (Fig.~3) are comparable to $\alpha=25^\circ$ (Fig.~2(f),(g)), higher harmonics are stronger developed for higher total magnetic fields. The observed dependence on the in-plane component is hinting towards an anisotropic spin scattering, which was recently predicted in a theoretical study\cite{Anisotropy}.

In conclusion we have reported AB effect in monolayer graphene with an experimental setup that allowed to tilt the device in respect to magnetic field. A development of the first harmonic and the observation of the second harmonic were presented with an increase of the total magnetic field. This observation is explained by an increase of the phase coherence length through spin polarisation. Additionally we show that increasing the in-plane component alone leads to a similar result, hinting towards a further anisotropic decoherence mechanism.

We acknowledge discussions with V.~I.~Fal'ko, P.~Recher and N.~Ubbelohde. This work was supported by the DFG via SPP 1459, and the NTH School for Contacts in Nanosystems.

\end{document}